\documentclass{article}

\usepackage[utf8]{inputenc} 
\usepackage[T1]{fontenc}    
\usepackage{hyperref}       
\usepackage{url}            
\usepackage{amsfonts}       
\usepackage{nicefrac}       
\usepackage{microtype}      
\usepackage{lipsum}
\usepackage{fancyhdr}       
\usepackage{graphicx}       
\usepackage{amsmath,amsthm,amssymb}

\newtheorem{thm}{Theorem}

\newtheorem{cor}{Corollary}
\newtheorem*{defn}{Definition}
\newtheorem*{rem}{Remark}

\graphicspath{{Figures/}}     

\title{Disenfranchisement in Spatial Voting Games
}

\author{
  Vincent Lageschulte, Joel Nishimura \\
  School of Mathematical and Natural Sciences \\
  Arizona State University \\
  Glendale, AZ\\
}

\begin{document}
\maketitle

\begin{abstract}
The questions of voting rights, group membership, and the boundaries of acceptable discourse all involve a recursive, self-stability issue, wherein the voters, group members, and acceptable commentators are often the ultimate arbiters of their own rules for inclusion and exclusion. In order to understand how ideological or issue diversity may impact the stability of such inclusion and exclusion, we introduce franchise elections, where voting members of a group vote to change the electorate. In particular, when agents are positioned inside an ideological or issue space, a social choice function that aggregates agent viewpoints induces a preference for agents over group composition, where agents prefer groups that improve the social choice function outcome for them. We demonstrate that when ideological space is two-dimensional, these franchise elections can generate complex behavior and that this complexity exists for any social choice function. Using the Yolk social choice function from voting theory, we illustrate that franchise elections can predict that agents attempting to introduce a new ideological axis perpendicular to an existing left-right spectrum risk disenfranchisement unless they ally with either side of the spectrum, and that franchise elections can cycle with any period. 

\end{abstract}

\section{Introduction}

 In a polity, who gets to vote? In a discourse, who gets to participate? Or in a social group, who is included? 
 Membership is an essential question for any group, as membership ultimately defines the behaviors and policies of the group, often including future decisions about expanding or restricting membership. If membership determines group decisions, and members have preferences over those group decisions, then group decisions induce preferences over group membership. 

The feedback between group membership, group social choice, and preferences for changing group membership has motivated a rich and diverse body of research that combines several disparate areas of research. The question of how a group, once constituted, can aggregate opinions into a single overall decision is explored in the social choice literature and its intersections with voting theory. The question of how and when groups change their membership has been pursued as both a vibrant historical question, mostly applied to early pre-modern Western countries, and a theoretical question. 

The question of what social choice functions allow for group consistency in regards to members unilaterally choosing to leave and other design criteria has spawned a productive line of research in \cite{barbera2001voting,berga2004stability,bergantinos2017societies,bergantinos2018societies}.

The historically motivated question of why a nation's elite would expand the franchise has been framed as a choice to preemptively prevent revolt \cite{acemoglu2000did, conley2001endogenous}, a question of intra-elite bargaining positions for public goods \cite{lizzeri2004did}, and intra-elite positioning in relation to industrialization \cite{llavador2005partisan}. In these works, the ideological positioning of the elite is viewed largely along a one dimensional axis, and additional modeling assumptions are drawn from the historical record to motivate the expansion of the electorate. 

An attempt to provide an overarching theory of the stability of such systems was provided in Acemoglu, Egorov and Sonin \cite{acemoglu2012dynamics}. Acemoglu, Egorov and Sonin developed an impressive framework for describing the eventual outcome for farsighted agents provided the population does not have innate cycles in their preferences. 

The question of group self-consistency also arises in the question of the structure of discourses. Named after John Overton, the Overton window is a framework designed to help explain what policies are viewed as politically active, and which are viewed as unacceptably extreme \cite{lynch2020regimes}. Borrowing from the more colloquial and expanded version of this concept, we will use the Overton window to refer to the set of viewpoints regarded as socially acceptable to espouse in a discourse. Principally, one feature of that Overton window is self-consistency, where people with views inside the Overton window have the most leverage in deciding which views are acceptable, and thus within, the Overton window. 

While we may retrospectively imagine that historical eras had only a few defining narrow political issues, so that they are adequately described by a one-dimensional issue space as in Acemoglu, Egorov and Sonin, discourses seem more likely to be messy and mix many different and wide ranging issues, and agents in a discourse can express opinions that cannot be neatly placed in a one-dimensional opinion space. This suggests that the question of group self-consistency may benefit from another look, where issue space allows for more complexity. Thus, we will base our analysis of group self-consistency on phenomena that can occur in an opinion space of two and greater dimensions.

This paper operates in the setting where cycles in group preferences are possible. We propose a general framework of franchise and disenfranchisement in section 2, where the a social choice function induces voter preferences over voting subgroups. In section 3 we illustrate this framework using the well known social choice function of the Yolk, and demonstrate the dangers of attempting to introduce new political issues while remaining neutral on existing issues. In section 4, we prove that there are populations that are disenfranchisement unstable regardless for all reasonable social choice functions. In section 5, we show that this framework can describe settings where individuals even elect to disenfranchise only themselves. In section 6, we describe the franchise graph, that governs the potential general dynamics of changes to the franchise and show that that graph can contain subgraphs that are cycles and cliques of arbitrary size.

Indeed, we show that a relatively simple and well known setting, Euclidean preferences with spatial voting, can produce not just cyclic group preferences in two Euclidean Dimensions, but cycles of arbitrary length. Group membership, or franchises can thus be unstable even under relatively benign modeling assumptions. Further, we show that issues of group instability are fundamental regardless of the choice of social choice function. As an example, we model the Overton Window of public opinion when there is an established left-right linear embedding and a concerned group wishes to introduce a new axis in the dialogue. In that setting, we show the limits of neutrality.

\section{Franchise Elections}

Consider an ideological space $V$ with metric $d$ and a population of voter ideal points $N$, which is a multiset of $V$ of size $n$. Voters have utilities $u:V\times V \to \mathbb{R}$, where $u(x,p)$ denotes the utility that a voter with ideal point $x$ has for proposal $p$. We will often assume that $u(x,p) = -d(x,p)$ so that voters prefer closer ideal points. Given a voting population $A\subset N$ of size $m$, let the social choice function $f:V^m \to V$, represent the current group choice\footnote{This social choice function may represent the average of many elections or simply represent some arbitrary mapping from voter preferences to a choice.}. We thus assume that the expected utility that a voter with ideal point $x$ receives from the current voting population is $u(x,f(A))$. With this framework, and for a fixed social choice function $f$, voters can compare their utilities between different subpopulations, and thus may prefer enfranchising one subpopulation over another.

We consider a franchise election as a vote on a proposal for the voting population $A$ to change to a new voting population $A'$. When $A'\subset A$ we refer to the election as a disenfranchisement election. When the population exists in a vector space, a specific kind of franchise election is an Overton-franchise election. In an Overton franchise election, we assume that the status quo and proposed subpopulations include all members of the population within their respective convex hulls. Namely, in an Overton franchise election there is a current convex Overton window $K\subseteq V$ and a proposal convex Overton window $K'\subseteq V$ and the status quo subpopulation is $A = N\cap K$, and the proposed subpopulation is $A' = N \cap K'$. 

Franchise elections in which the winners are determined by majority vote will be the primary focus of this paper. 
Throughout this paper, we assume that voters are myopic, so voters with ideal point $x$ will vote for the status quo if $u(x,f(N))>u(x,f(M))$, will vote for change if $u(x,f(N))<u(x,f(M))$, and otherwise will abstain. 

In our judgment, the scenario of an election that changes an electorate by strictly decreasing the number of the voters, merits special attention. To enable a more clear distinction between changes to the electorate generally, and changes which strictly decrease the electorate, consider the following two definitions:

\begin{defn}[Franchise Stability] 
    We say that a population $N$ is franchise unstable if there exists a population $M$ such that $M$ wins a franchise election against $N$.  If $N$ is not franchise unstable, then we say it is franchise stable.
\end{defn}

\begin{defn}[Overton Stability]
   We say that a population $N$ which contains all voter ideal points in a convex Overton window $K$ is Overton unstable if there exists an Overton window $K'$ and population $M$ containing all voter ideal points in $K'$ such that $M$ wins a franchise election against $N$.  If $N$ is not Overton unstable, we say it is Overton stable.
\end{defn}

\begin{defn}[Disenfranchisement Stability] 
    We say that a population $N$ is disenfranchisement unstable if there exists $M\subset N$ such that $M$ wins a franchise election against $N$.  If $N$ is not disenfranchisement unstable, then we say it is disenfranchisement stable.
\end{defn}

Notice that a franchise stable subpopulation is also Overton and disenfranchisement stable, but Overton and disenfranchisment stability do not imply each other.
Moreover, these concepts are more specific than the notion of a core in standard spatial voting, which considers challenging the status quo against all potential proposals in $V$, not just those that can be expressed as $f(M)$ for some $M$.

\begin{thm}
    If population $N$ has a core $c$ and $f(N)=c$, then it is franchise stable.
\end{thm}
\begin{proof}
    Since no point in $V$ is strictly preferred to $c$, there does not exist a different population $M$ such that $f(M)$ is strictly preferred to $f(N)$. 
\end{proof}
As a corollary, we apply Black's median voter theorem \cite{black1948rationale} to conclude that 1-D embeddings with a median social choice function are always franchise stable. 
\begin{cor}
    For majority elections on $V =\mathbb{R}$, if the social choice function selects the median of population $N$, then $N$ is franchise stable.  
\end{cor}

Since franchise, and disenfrachisement elections restrict the set of possible competing proposals, more scenarios are stable here than in standard spatial voting.

\section{Yolk-Disenfranchisement}

To construct an explicit example, we will choose an ideal space and a social choice function defined on that space. For this section, let the ideal space be $V = \mathbb{R}^k$ with the Euclidean metric. For the social choice function, we will consider the well-established yolk first introduced by \cite{mckelvey1986covering}, which we review below. 

The yolk is properly defined after introducing median hyperplanes. For a population or subpopulation $A$ with $n$ voter ideal points in Euclidean space, a median hyperplane is a hyperplane that divides space into two closed half spaces $H^+$ and $H^-$ so that both half spaces contain at least half of the population: $|H^+\cap N|\ge\frac{n}{2}$ and $|H^-\cap N|\ge\frac{n}{2}$.
\begin{defn}[Yolk]
    In Euclidean space, the population or subpopulation $A$ has a yolk with center $c$ and radius $r$ if the ball with radius $r$ centered at $c$ is the smallest ball that intersects every median hyperplane.
\end{defn}
The yolk is well known to have many interesting properties, and to also be close to similar concepts, like the finagle point \cite{nganmeni2023finagle}. For instance, proposals distant from the yolk center are always beaten by some proposal within a multiple of the yolk radius \cite{feld1989limits}. 

Taken together, these properties of the yolk suggest that its location is a capable summary---although not necessarily uniquely capable---of potential outcomes from electoral games played with population $P$. Indeed, as argued from several different viewpoints, there cannot be a perfect summary of electoral games, partly because there cannot be any perfect social choice function on Euclidean space. In some situations, the yolk is not uniquely defined \cite{martin2016uniqueness}, but since the set of possible yolk centers is convex, let $y(A)$ denote the center of mass of possible yolk centers, and then let our social choice function be $f(A) = y(A)$.   

Now, we will consider an explicit example of the disenfranchisement game with ideal space $V = \mathbb{R}^2$ and social choice function $y(A)$. 

\subsection{A Line and an Iconoclast}
Consider a population that has largely sorted itself into a familiar left-right political spectrum and a single iconoclast group that is trying to introduce a new issue, represented by a secondary ideological axis. For example, suppose that a scientist or a group of scientists is seeking to motivate action on an urgent issue in a society that has no prior commitments on it. To model this, suppose population $N=M\cup {A}$ where $M$ is a subpopulation entirely on the x-axis with median $\mathbf{0}$, and $A$ is an iconoclast or group of iconoclasts off the x-axis. 

\begin{figure}
    \centering
    \includegraphics[width=1.0\linewidth]{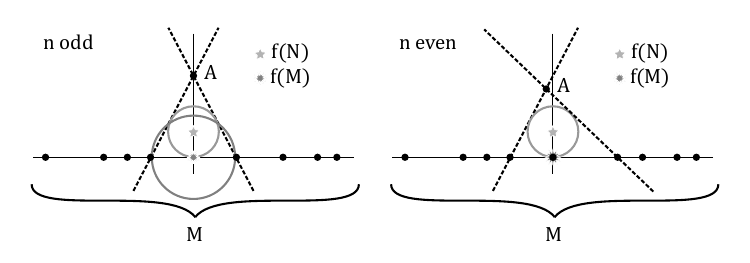}
    \caption{Population $N$ has voters located at the black points. The extremal median lines form a triangle with $A$ at its apex and this moves the yolk (grey circle) above the x-axis. Since $f(M)$ is collinear with $M$ and $f(N)$ lies strictly vertical of $f(M)$ every voter in $M$ is closer to $f(M)$ than $f(N)$ and thus $N$ is disenfranchisement unstable for any election where $A$ does not have a unilateral veto.  }
    \label{fig:y_axis_deviant}
\end{figure}

It may seem like $A$'s optimal course of action to introduce and retain another issue axis among the overall population is to take an ideal stance close to the median, which would avoid the controversy associated with positions further left or right. However, this is not the case, as in figure~\ref{fig:y_axis_deviant}, if $f(N)$ lies strictly vertical of $\mathbf{0}$, then $N$ is necessarily disenfranchisement unstable for any election where $A$ does not have a unilateral veto because $N$ loses in an election against $M$. To relate this to the earlier example, it seems that by introducing the urgent issue without allying with either side, neither side has enough incentive to wish to keep the scientists and their iconoclastic stance in the bounds of mainstream political discourse. Hence, iconoclasts with a stance on the primary axis close to the median of the rest of the population are not likely to remain in the Overton window.

In contrast, if $A$ were to move its position sufficiently towards either political pole, $A$ would be able to ally with one political side, and this alliance would provide the political cover necessary for $A$ to remain within acceptable discourse, as we show below.

\begin{figure}
    \centering
    \includegraphics[width=1.0 \linewidth]{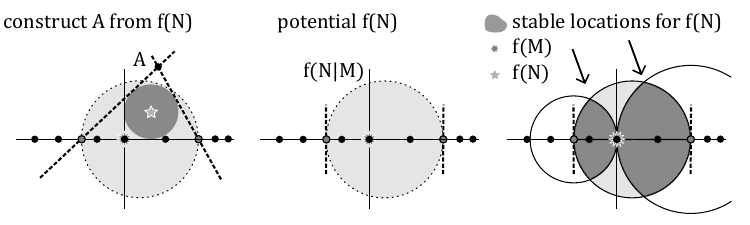}
    \caption{Here $A$ represents a group of $3$ voters and thus there are three members of $M$ between the median hyperplanes through $A$. For a potential center $f(N)$, the point $A$ that supports that center can be constructed as on the left. The set of $f(N)$ for all positions of $A$ is a circle that places the pivotal voters opposite of each other (center). The location of a stable $f(N)$ must also be closer to at least one pivotal voter than $f(M)$ (right).    }
    \label{fig:optimal_camoflage}
\end{figure}

Suppose that $A$ represents a group of $m$ voters, and let $\mathbf{x}_l$ and $\mathbf{x}_r$ denote the pivotal voters on the left and right so that planes through $A$ and each of $\mathbf{x}_l$ and $\mathbf{x}_r$ are both median planes. Say that $A$ wishes to join the population at an ideal point that will move the population's choice to some chosen $f(N)$, which $A$ wants to be as far from the x-axis as possible. To keep the new issue axis in the voting subpopulation, $A$ must choose $f(N)$ so that any subpopulation of $N$ that wins a franchise vote against $N$ contains $A$. Hence, $A$ will consider two competing forces---the distance between $f(N)$ and its closest pivotal voter compared to the distance between $f(M)$ and that pivotal voter, and the distance between $f(N)$ and $f(M)$. For $A$ to remain in the voting subpopulation, $f(N)$ must be closer to the pivotal left or right voter than $f(M)$; otherwise, more than half of the voters prefer $f(M)$ to $f(N)$. Additionally, since $A$ only represents $m$ votes, $A$ is limited in its ability to move $f(N)$ far from $f(M)$, so $f(N)$ must be sufficiently close to $f(M)$. These counterbalancing forces, as seen in figure~\ref{fig:optimal_camoflage}, give that $A$'s best choices for $f(N)$ lie at the intersections of two circles, where one intersection is the choice for $f(N)$ if $A$ allies with the left, and the other is the choice if $A$ allies with the right. Once the location for $f(N)$ is set, $A$ chooses a corresponding ideal point using the yolk centered at $f(N)$. Given that $|N|$ is odd, the median lines are the x-axis and two lines tangent to the yolk to the left and right of $f(N)$. Because the yolk is tangent to at least three median lines, the yolk is tangent to all of $N$'s median lines. Thus, the yolk centered at $f(N)$ is tangent to the x-axis, and to correctly place the other two median lines, $A$ chooses the intersection of the lines through the pivotal voters tangent to the yolk, as in figure~\ref{fig:optimal_camoflage}. This construction fails if the lines through the pivotal voters are parallel, and a simple trigonometric argument reveals that this occurs when $f(N)$ lies on the circle centered on the x-axis that contains the two pivotal voters. 

Interestingly, if the sole goal of $A$ is to position itself to maximize the impact of the new issue, as measured by the height of the y-axis of the stable value of $f(N)$, then $A$ should actually position itself on the side whose pivotal voter is farther from $f(M)$. In other words, and perhaps counter-intuitively, $A$ should move towards the more extreme of the two pivotal voters. 

However, the introduction of $A$ can be destabilizing even if $A$ is not vulnerable to being voted out of the Overton window. For linear embeddings, a consequence of the median voter theorem is that the far left and right voters will not be ejected for their viewpoints. However, the presence of $A$ off the line creates a tension that may lead to disenfranchising voters on the extremes. While $\mathbf{x}_l$ and $\mathbf{x}_r$ are the pivotal voters on the median lines through $A$ on the left and right respectively, let ${\mathbf{x}_l}^*$ and ${\mathbf{x}_r}^*$ represent the winning voters on the left and right, so that ${\mathbf{x}_l}^*$ is the leftmost voter such that the line through it and $A$ encloses a winning majority on the left, and similarly for ${\mathbf{x}_r}^*$ on the right. When $n$ is odd, the pivotal voter and the winning voter are the same.

\begin{figure}
    \centering
    \includegraphics[width=1.0\linewidth]{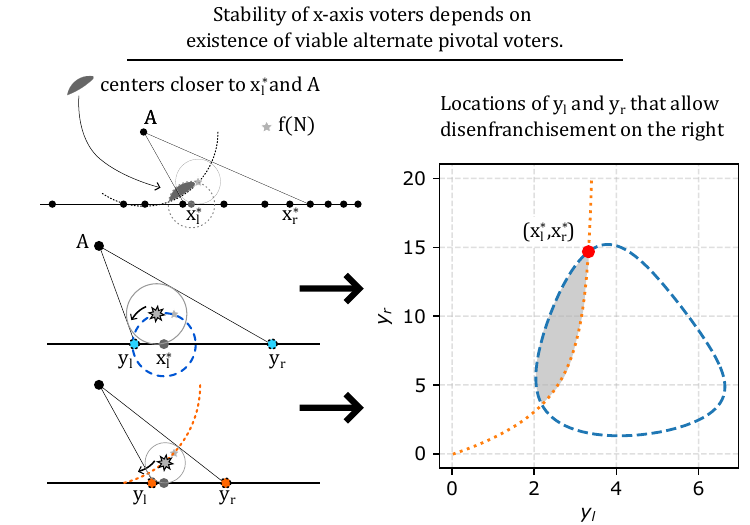}
    \caption{Here $A$ represents a group of $6$ voters with a total of $18$ total voters, this leads the winning voters ${\mathbf{x}_l}^*$ and the ${\mathbf{x}_r}^*$ to lie between the $\mathbf{x}_l$ and $\mathbf{x}_r$. A proposal to disenfranchise a voter or multiple voters on the right will pass if both ${\mathbf{x}_l}^*$ and $A$ support it, which requires that the new center lie within shaded region of the upper left plot, which is within the circles centered at ${\mathbf{x}_l}^*$ and $A$ through the current center. In order for the new center to lie in that shaded region there must be voters capable of being the new pivotal voters (so that the new center is the inscribed triangle between the vertices at $A$ and the new pivotal voters). For the current configuration, if there are viable new pivotal voters in the shaded region (right plot) then the system is disenfranchisement unstable.  }
    \label{fig:vote_out_extremes}
\end{figure}

As seen in figure~\ref{fig:vote_out_extremes}, there are situations where a majority votes for a change in the electorate that disenfranchises voters on the line. Without loss of generality, suppose $R$ is any member of the population to the right of $x_r$. Eliminating $R$ from $N$ can result in all of the pivotal and winning voters moving left. When $\mathbf{x}_l$ and/or $\mathbf{x}_r$ move left, the new yolk center, $f(N\setminus R)$ also moves left. To understand the stability of this situation, let $p=|R|$, and let $\mathbf{y}_l(p)$ and $\mathbf{y}_r(p)$ denote sequences of the $p$ points to the left of $\mathbf{x}_l$ and $\mathbf{x}_r$ respectively, so that $\mathbf{y}_l(1)$ is the first voter to the left of $\mathbf{x}_l$ and $\mathbf{y}_l(2)$ is second, and similarly for $\mathbf{y}_r$ and $\mathbf{x}_r$. Thus, this sequence contains the potential alternate pivotal voters depending on the number of disenfranchised voters on the right. Disenfranchisement stability requires that no pair of $(\mathbf{y}_l(p),\mathbf{y}_r(p))$ produces a center preferable to both $A$ and ${\mathbf{x}_l}^*$. As in figure~\ref{fig:vote_out_extremes}, the locations of voters that lead to this disenfranchisement instability can be plotted as the region enclosed by two curves, each of which is found by locating the yolk center along the boundary of circles centered at ${\mathbf{x}_l}^*$ and $A$ respectively. Hence, stability requires that the sequence $(\mathbf{y}_l(p),\mathbf{y}_r(p))$ is disjoint from this region. Thus, the disenfranchisement stability of the line and a single group $A$ off the line depends subtly both on the location of the $A$ and the sequence of alternate possible pivotal voters on the line. 

Since the only possible changes to the convex Overton window are the elimination of $A$ and extreme voters on either the left or right, this scenario is Overton stable provided that $A$ is located in the intersection of the two circles, as in figure~\ref{fig:optimal_camoflage}, and the set of alternate left pivotal voters avoids the regions illustrated in figure~\ref{fig:vote_out_extremes} and similarly for the set of alternate right pivotal voters. 

This analysis demonstrates that, while it is possible for a group to introduce a second ideological axis to a linear polity and maintain franchise stability, they must ally with either the left or right. Furthermore, franchise stability is also subject to subtle conditions on the locations of alternate pivotal voters on the line. However, the iconoclast group is not necessarily concerned with this latter consideration for franchise stability, as they are not immediately and possibly never at risk of being disenfranchised due to extreme left or right voters being disenfranchised.

\section{Instability for General Social Choice Functions}

Although we have shown that some populations are franchise unstable when paired with particular social choice functions, thus far we have left open the possibility that there exists a social choice function that, like the median in 1-D, results in franchise stability for all populations.
However, as we will show in this section, there does not exist such a social choice function. We will demonstrate that in any Euclidean space of two or more dimensions there exists a franchise unstable population for all reasonable social choice functions. 

To restrict the set of social choice functions to reasonable options, we will only consider those with the unanimity property, which is defined as follows. First, we say that a population $N$ has a unaminous consensus $c$ if $x=c$ for all $x\in N$. Next, let us define the following condition on a social choice function.
\begin{defn}[Unanimity]
    The social choice function $f$ has the unanimity property if for all populations $N$ with unanimous consensuses $c$, $f(N)=c$.
\end{defn}
All weakly Pareto-efficient social choice functions have this property, so given that social choice functions without weak Pareto efficiency would not provide reasonable summaries of group choice in this context, we can say that reasonable social choice functions have this property.

While this condition seems weak, it can lead to franchise instability in some populations. To demonstrate this, first, we define a total dominating set, which is a variation of a dominating set that also requires that it dominates its own members.
\begin{defn}[Total dominating set]
    For population $N$, a set of proposals $P$ is a total dominating set for $N$ if for any proposal $x\in V$ there exists $p\in P$ so that $p \succ x$, where $p\neq x$.
\end{defn}

Note that the union of two disjoint dominating sets is necessarily a total dominating set. Also, because no point dominates a core, any population that has a core cannot have a total dominating set. Thus, total dominating sets exist only for populations without cores. Total dominating sets can also be used to construct franchise unstable populations for any social choice function with the unanimity property. 
\begin{thm}
    For any population $N$ with $n = |N|$ and a total dominating set $D$ with $k=|D|$, and for any social choice function $f$ with the unanimity property; there exists a franchise unstable population $M$ with $(k+1)n+k$ voters located only at points in $N$ and $D$.
    \label{thm:total_dom_franchise}
\end{thm}
\begin{proof}
    Construct population $M$ in the following manner. First, place $k+1$ copies of each member of $N$ in population $M$. Next, place one copy of each of the set $D$ into $M$. Now consider any winning majority of population $N$, and notice that the same locations in $M$ constitute a majority, since those locations now represent at least $(k+1)\lceil n/2 \rceil$ which is a strict majority of the total population, $(k+1)n +k$ votes. Thus, the population $M$ will decide competitions between proposals exactly the same as $N$, and $D$ will also be a total dominating set of $M$. Suppose $f(M)=c$, since $D$ is a total dominating set, there exists some $d\in D$ such that $d\succ c$, and thus $f(\{d\})\succ f(M)$ implying that the proposal to disenfranchise all members of $M$ except for $d$ beats the status quo in a majoritarian election. This shows that $M$ is franchise unstable.
\end{proof}

Notice that this theorem does not require that voters be embedded in Euclidean space, and as shown in the appendix, this theorem is true if voter preferences depend only on the voter's and the proposal's location in some space.

\begin{figure}
    \centering
    \includegraphics[width=1.0 \linewidth]{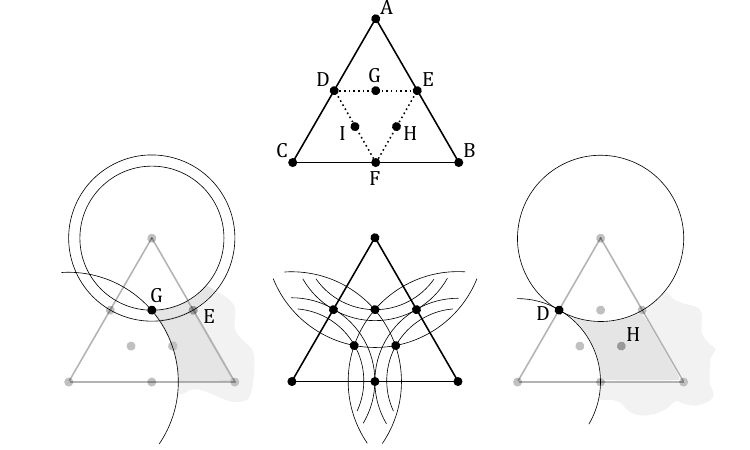}
    \caption{The points $D-I$ are a total dominating set for both $N= \{A,B,C\}$ and a population $M$ that has one voter at points $D-I$ and $7$ voters at each of $A$, $B$ and $C$. That $D-I$ are a total dominating set for $N$ can be visually inspecting circles centered at the corners (middle bottom), for instance the region that $G$ dominates is shaded on the left, and that $D$ dominates is shaded on the right.  }
    \label{fig:Triangle27}
\end{figure}

\begin{cor}
For $V=\mathbb{R}^k$ for $k\ge 2$ and any social choice function $f$ with the unanimity property, there exists a population $N$ that is franchise, Overton and disenfranchisement unstable. \label{cor:some_N_unstable}
\end{cor}
\begin{proof}
    By theorem~\ref{thm:total_dom_franchise}, it suffices to produce a population with a total dominating set. Consider, as in figure \ref{fig:Triangle27}, an equilateral triangle. These voters have a total dominating set composed of the six points in figure \ref{fig:Triangle27}, where points $D$, $E$, and $F$ lie on the midpoints of the triangle's sides, and points $G$, $H$ and $I$ are on the midpoints of $\bar{DE}$, $\bar{EF}$ and $\bar{DF}$ respectively. $D$, $E$ and $F$ clearly dominate all other points in the space, since any point must be farther than half the side length from at least two of the three triangle vertices. However, $D$, $E$ and $F$ don't dominate each other. Instead, $D$, $E$ and $F$ are each dominated by the two closer points of $G$, $H$ and $I$. Thus, by theorem~\ref{thm:total_dom_franchise}, there exists a population of $27$ voters with $21$ voters at the corners and one voter at each of $D-I$ that is franchise unstable for any $f$ with the unanimity property. Since the instability is due to a proposal that selects only a single voter inside the triangle, this franchise instability also implies an Overton and disenfranchisement instability.

\end{proof}

Since Pareto optimal social choice functions have unanimity, this also implies that there are populations that are franchise unstable for all Pareto optimal social choice functions. 

Unanimity is necessary for the above result because there exist social choice functions without unanimity that can create stability. For example, the constant social choice function $g(N)=0$ for all populations $N$ is stable for all populations. Indeed, since a constant social choice not depend on the choice of subpopulation there does not exist a possible change in the voting population that improves any voters' utilities, so no proposal to change the enfranchised population can pass, and any population is franchise stable.

Moreover, beyond the constant social choice functions that are, in a way, neutral to disenfranchisement, given any fixed population $N$, there also exists a somewhat trivial social choice function without unanimity that punishes disenfranchisement. For a fixed population $N$ such that $n=|N|$, with bound $B$ so that $B>||\mathbf{x}||$ for all $\mathbf{x}\in N$, consider a social choice function $f_p$ that behaves as follows: $f_p(N)$ is some point in the convex hull of $N$, and for all $M\subset N$, $f_p(M)$ is some point at a distance from the origin greater than $3B$. Since any disenfranchisement results in $f_p$ lying at least $2B$ away from all members of $N$, while the status quo is, at most, $2B$ away, $f_p$ stabilizes $N$. Although, this method of stabilization is more like a crude threat against disenfranchisement than a true expression of social choice. Outside of ensuring that groups of students include all of their assigned group members in a class project, it is unclear how frequently or appropriately such a social choice function could be enforced, or whether this kind of social choice could arise and persist naturally.  

\section{Self-Disenfranchisement}

The negative results in corollary~\ref{cor:some_N_unstable} are built on a scenario where, given that the social choice function has the unanimity property, a majority of the population votes to disenfranchise themselves to empower a single member of the population whose position they prefer to the status quo. Although this situation may be counterintuitive, the self-disenfranchisement it presents is apparently reasonable in that voters elect to disenfranchise all but one of the competing viewpoints, which can enfranchise a viewpoint closer to the voter than the status quo.

\begin{figure}
    \centering
    \includegraphics[width=1.0 \linewidth]{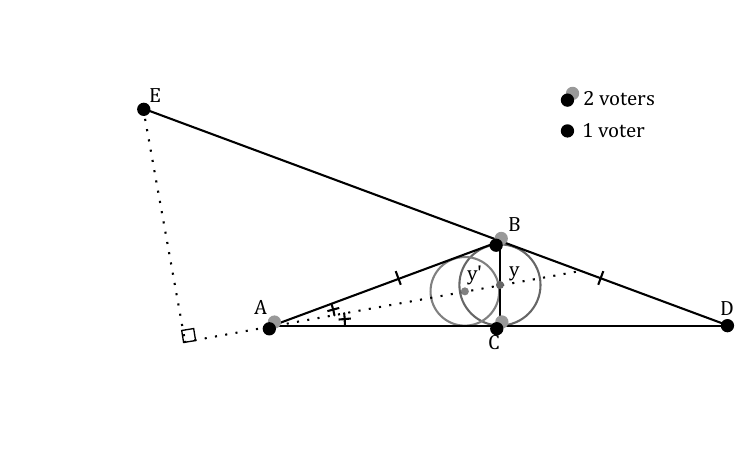}
    \caption{Population $N = \{A,A,B,B,C,C,D,E\}$ has the property that $y$, the yolk of $N$, is farther from $E$ than the yolk $y'$ of the population $M = N\setminus\{E\}$. Thus, $E$ may wish to vote themselves out of the electorate. }
    \label{fig:self_vote_out}
\end{figure}

Less intuitively, a voter may prefer to disenfranchise only themself, as in the following scenario. Consider the population $N$ in figure~\ref{fig:self_vote_out} with the yolk center of mass social choice function. When voter $E$ is present, the binding median half-planes for $N$ are the sides of the isosceles triangle $ABD$, yet when $E$ is absent, the binding median half-planes form the triangle $ABC$, which moves the new yolk along the line that bisects $A$ towards $E$. Thus, $E$ prefers not to be in the voting population. Hence, not only would $E$ prefer not to be truthful, 
but if $E$ must be truthful, then it would prefer not to be included at all. Although in this situation $E$ prefers to leave, the proposal to eliminate $E$ would not pass, as $B$, $C$, and $D$ would suffer by excluding $E$. In addition, voter $D$, the voter most distant from $E$, benefits surprisingly strongly from the inclusion of $E$. However, even if a majority vote to disenfranchise $E$ fails, in other situations, especially in real elections, voters may be able to enact self-disenfranchisement unilaterally. For example, while some countries enforce compulsory voting, many jurisdictions allow voters to abstain. 

From another perspective, because $E$ prefers to self-disenfranchise, $E$'s presence moves the center of the yolk farther away, but between any pair of proposals, $E$'s vote would only ever be determining in favor of a proposal preferred to $E$. On the one hand, this apparent paradox suggests that, while the yolk has many theoretical benefits, the center of the yolk is not always an accurate summary of the likely utility of a population to an ideal point. On the other hand, depending on the order in which agenda items are proposed, the long-term average of the winning proposals may be more likely to reside closer to the yolk's center, which suggests that, sometimes, when $E$ casts a decisive vote, it leads to an unviable status quo, vulnerable to competing proposals that are less favorable to $E$. Regardless, this scenario suggests that examining self-efficacy in multi-dimensional voting spaces through a game-theoretic perspective may provide a rational mechanism to describe observed low voter turnout, albeit in somewhat particular cases. 

\section{Franchise Dynamics}

While our investigation so far has been focused on the populations that may be enfranchised or disenfranchised given a current enfranchised population---or in other words, what may occur in a single election in the disenfranchisement game---this section focuses on the natural next question: what can happen if the disenfranchisement game continues? If we assume that, for an instance of the disenfranchisement game, there is one finite population of which all possible electorates are subpopulations, then there are a finite number of possible electorates. Thus, we can represent these electorates as nodes on a directed graph to demonstrate the dynamics of franchise elections.

\begin{defn}[Franchise Graph]

Let $N$ be a population of voter ideal points, and let $\mathcal{A}=\{A_1,A_2,...,A_n\}$ be the set of every non-empty subpopulation of $N$. Then the franchise graph $G$ of $N$ is the digraph with vertices $\mathcal{A}$ and arcs from $A_i$ to $A_j$ if a majority of the subpopulation of $A_i$ would vote to change to $A_j$ in a franchise election.

\end{defn}

If we know the franchise graph of a population, then we can infer the stability of it and all of its subpopulations from the properties of the franchise graph. For instance, a subpopulation $A$ is franchise stable if $A$ has no outgoing edges, disenfranchisement stable if there are no outgoing edges to subpopulations of $A$, and similarly Overton stable if there are no outgoing edges to subpopulations of $A$ expressible as Overton subpopulations. 

For any social choice function with the unanimity property, all populations have such stable points, since single point subpopulations are necessarily unanimous and thus stable. Hence, the question of whether stable points exist in the franchise graph $G$ is settled for all reasonable social choice functions and for any choice of franchise, Overton or disenfranchise stability. However, the question of whether there are any unstable configurations is partly answered in the negative for $V=\mathbb{R}$ by the median voter theorem, and in the affirmative for $V=\mathbb{R}^2$ by corollary~\ref{cor:some_N_unstable}. 

Since stable subpopulations exist, we next address the question of whether there may be cycles in $G$, and thus whether transitions on $G$ must eventually arrive at such stable points in a finite number of elections. To address whether cycles are possible, and to illustrate what cycles would look like, we will investigate using rotationally symmetric and anonymous social choice functions on Euclidean spaces.

\begin{thm}
Under any rotationally symmetric and anonymous social choice function $f$ in $\mathbb{R}^2$, there exists a population $N$ with a two-cycle in its franchise graph.
\label{thm:2D-2-cycle}
\end{thm}

\begin{proof}
    We will construct a population $N$ that has a two-cycle in its franchise graph. Let $A=\{a_1,a_2,a_3\}$ be a subpopulation of $N$ such that $a_1 = [1,2\sqrt{3}] $, $a_2= [1,-2\sqrt{3}]$, and $a_3 = [-5,0] $ lie on the vertices of an equilateral triangle. Then, because $f$ is rotationally symmetric and anonymous, $f(A)$ lies on the center of that equilateral triangle at $[-1,0]$. Now, let $B=\{b_1,b_2,b_3\}$ be a subpopulation of $N$ that is a reflection of $A$ across the $y$-axis such that the line segment from $b_1 = [-1,2\sqrt{3}]$ to $b_2=[-1,-2\sqrt{3}]$ is parallel to the line segment from $a_1$ to $a_2$ and intersects $f(A)$, and $b_1$, $b_2$, and $b_3$ are at the vertices of an equilateral triangle with its center on the line segment from $a_1$ to $a_2$. Then, a majority of $A$ favor the convex hull containing $B$ and excluding $A$, and a majority of $B$ favor the convex hull containing $A$ and excluding $B$. Therefore, there exists a two-cycle between subpopulation $A$ and subpopulation $B$.
\end{proof}

Notice that in the proof of Theorem 2, the two Overton subpopulations, $A$ and $B$ are disjoint. Thus, the construction in the proof of Theorem 2 continues to work even if the population of group $A$ and $B$ are scaled separately. For instance, if $A$ had 3 voters and $B$ had 6, the analysis in the above proof does not change. Thus, cycles in $G$ can involve both increases and decreases in subpopulation size, precluding any simple theorem about flows in $G$ tracking subpopulation sizes. 

Also, note that for myopic voters, this cycle is favored over $A\cup B$, but for farsighted voters with a sufficiently high discount factor, this cycle will not be favored over $A\cup B$. This suggests that cycles may occur even when voters have some farsightedness, but there may also be a lower bound for farsightedness that is sufficient to cause a cycle to lose to the union of all the subpopulations in the cycle in a franchise sequence election.

Next, we will demonstrate the possibility of longer cycles and complete graphs as subgraphs of franchise graphs. The following result uses the yolk social choice function, but the later results use any rotationally symmetric and anonymous social choice function.

\begin{thm}
    In $\mathbb{R}^2$ with the yolk center social choice function there exists a population $N$ with a three-cycle as a subgraph of its franchise graph.
    \label{thm:2D-3-cycle}
\end{thm}

\begin{figure}
    \centering
    \includegraphics[scale=0.3]{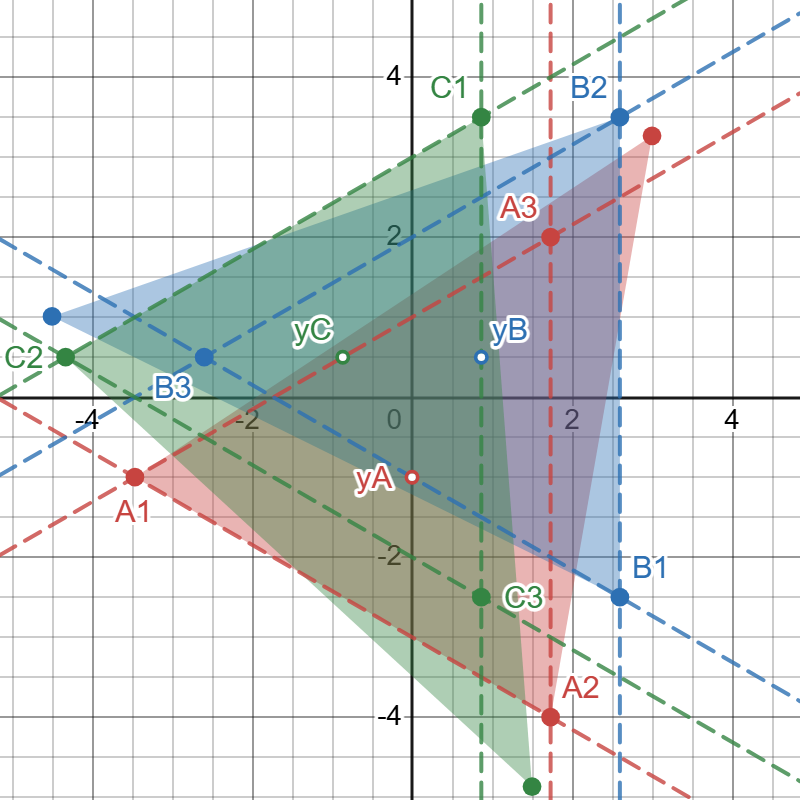}
    \begin{tabular}{l|llll}
                    & Vertex 1                                     & Vertex 2                                     & Vertex 3                                      & Vertex 4                                         \\ \hline
    Subpopulation A & $4\times (-2\sqrt{3},-1)$                    & $4\times (\sqrt{3},-4)$                      & $2 \times (\sqrt{3},2)$                       & $1\times(3,5-\sqrt{3})$                          \\
    Subpopulation B & $4\times (\frac{3\sqrt{3}}{2},-\frac{5}{2})$ & $4\times (\frac{3\sqrt{3}}{2},\frac{7}{2})$  & $2 \times (-\frac{3\sqrt{3}}{2},\frac{1}{2})$ & $1\times (-\frac{9}{2},14-\frac{15\sqrt{3}}{2})$ \\
    Subpopulation C & $4 \times (\frac{\sqrt{3}}{2},\frac{7}{2})$  & $4\times (-\frac{5\sqrt{3}}{2},\frac{1}{2})$ & $2 \times (\frac{\sqrt{3}}{2},-\frac{5}{2})$  & $1 \times (\frac{3}{2},-4-\frac{\sqrt{3}}{2})$  
    \end{tabular}
    \caption{The population $N= A\cup B\cup C$ has a three-cycle, $(A,C,B)$ in its franchise graph. The convex hull of $A$ is shaded in red, $B$ in blue, and $C$ in green. Median lines that bound the yolks are noted in the subpopulations' respective colors.}
    \label{fig:2D-3Cycle}
\end{figure}

\begin{proof}
    Consider a population $N$ constructed and displayed as in Figure \ref{fig:2D-3Cycle}, and  consider three different convex enfranchisement rules that include the convex hulls of subpopulations $A$, $B$ and $C$. Each of those convex hulls forms a triangle, and when $A$ is entirely enfranchised it includes $C_3$, when $B$ is enfranchised it includes $A_3$ and when $C$ is totally enfranchised it includes $B_3$, so that the size of each subpopulation is $13$. Thus, the three bounding median lines for each subpopulation are those through vertices $1$, $2$, and $3$. These points are chosen so that the yolk centers, $y_A$, $y_B$ and $y_C$ are distributed regularly around the unit circle, with $y_A=(0,-1)$, $y_B=(\frac{\sqrt{3}}{2},\frac{1}{2})$ and $y_C=(-\frac{\sqrt{3}}{2},\frac{1}{2})$. Now consider the viewpoint of subpopulation $A$. If $y_A$ is the status quo, $A_1$, $A_3$, and $A_4$ are closer to $y_C$ than $y_A$, so $7$ of $13$ voters favor changing the electorate to $C$ rather than keeping $A$. Similarly, if $C$ is the status quo, then $C_1$, $C_3$, and $C_4$ are closer to $y_B$ than $y_C$, so a majority of $C$ favors $B$. Lastly, if $B$ is the status quo, then $B_1$ $B_3$, and $B_4$ are closer to $y_A$ than $y_B$, so a majority of $B$ favors $A$. Therefore, there is a three-cycle, $A\to C \to B$.

\end{proof}

Moving beyond $\mathbb{R}^2$ as the voter ideal space, more dimensions enable general constructions for populations whose franchise graphs feature $n$-cycles or $n$-complete graphs as subgraphs. For example, in $\mathbb{R}^3$ we can easily construct a population with an $n$-cycle as a subgraph for $n>2$, and in $\mathbb{R}^{n-1}$, we can construct a population with an $n$-complete graph as a subgraph.

\begin{thm}
    For $n\ge 2$ and any rotationally symmetric and anonymous social choice function $f$, there exists a population in $\mathbb{R}^3$ with an $n$-cycle in its franchise graph.
    \label{thm:3D-n-cycle}
\end{thm}

\begin{proof}
    We will construct a population $N$ with an $n$-cycle in its franchise graph. Let 
    \begin{eqnarray*}
    a_{k,1}&=&\biggl(3\cos\biggl(\frac{(2k+2)\pi}{n}\biggr)-2\cos\biggl(\frac{2k\pi}{n}\biggr),3\sin\biggl(\frac{(2k+2)\pi}{n}\biggr)-2\sin\biggl(\frac{2k\pi}{n}\biggr),0\biggr)\\
    a_{k,2}&=&\Biggl(\cos\biggl(\frac{2k\pi}{n}\biggr),\sin\biggl(\frac{2k\pi}{n}\biggr),\sqrt{3}\sqrt{2-2\cos\biggl(\frac{2\pi}{n}\biggr)}\Biggr)\\
    a_{k,3}&=&\Biggl(\cos\biggl(\frac{2k\pi}{n}\biggr),\sin\biggl(\frac{2k\pi}{n}\biggr), -\sqrt{3}\sqrt{2-2\cos\biggl(\frac{2\pi}{n}\biggr)}\Biggr)\\
    A_k &=&\{a_{k,1},a_{k,2},a_{k,3}\}
    \end{eqnarray*}
    for $k\in\{1,2,...,n\}$, so each $A_k$ contains the vertices of an equilateral triangle, and let $N=\bigcup_{k=1}^nA_k.$ Then, because $f$ is rotationally symmetric and anonymous, $f(A_k)=(\cos(\frac{(2k+2)\pi}{n}),\sin(\frac{(2k+2)\pi}{n}),0)$, the center of the triangle with vertices at $a_{k,1}$, $a_{k,2}$, and $a_{k,3}$. For $k>1$, $a_{k,2}$ and $a_{k,3}$ are closer to $f(A_{k-1})$ than $f(A_k)$, so a majority of $A_k$ prefer $A_{k-1}$. For $k=1$, $a_{1,2}$ and $a_{1,3}$ are closer to $f(A_n)=(\cos(\frac{(2n+2)\pi}{n}),\sin(\frac{(2n+2)\pi}{n}),0)=(\cos(\frac{2\pi}{n}),\sin(\frac{2\pi}{n}),0)$ than $f(A_1)=(\cos(\frac{4\pi}{n}),\sin(\frac{4\pi}{n}),0)$, so a majority of $A_1$ prefer $A_n$. Therefore, $N$ has an $n$-cycle, $A_n\to A_{n-1}\to\cdots\to A_1$, in its franchise graph.
\end{proof}

\begin{rem}
Following this construction for $n=2$ gives the population with a two-cycle described in the proof of \ref{thm:2D-2-cycle} on the xz-plane.
\end{rem}

\begin{figure}
    \centering
    \includegraphics[scale=0.3]{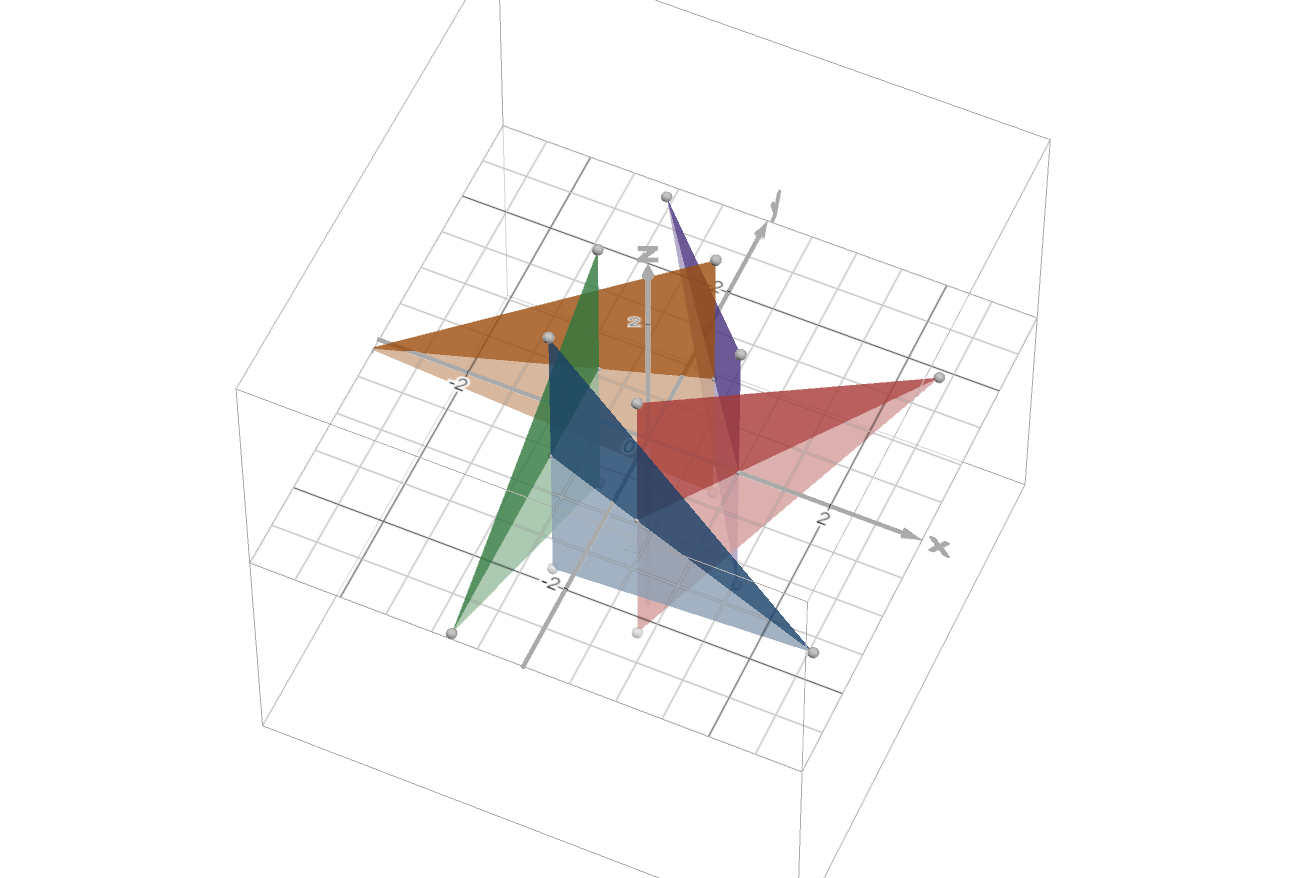}
    \caption{This is a population in $\mathbb{R}^3$ with a five-cycle in its franchise graph. The shaded triangles are the convex hulls of the subpopulations in the cycle, and the vertices of the triangles are the voter ideal points.}
    \label{fig:3D-5Cycle}
\end{figure}

Next, we show that cliques of up to $n+1$ populations can appear when the voter ideal space is $n$-dimensional, starting with a clique of 4 in $\mathbb{R}^3$.

\begin{thm}
For any rotationally symmetric and anonymous social choice function $f$, there exists a population in $\mathbb{R}^3$ whose franchise graph contains as a subgraph a 4-complete digraph.
\end{thm}

\begin{proof}
Let the following be subpopulations of population $N$:
    \begin{align*}
        A_1&=\left\{\left(5,4,4\right),\left(-4,-5,4\right),\left(-4,4,-5\right)\right\},\\
        A_2&=\left\{\left(4,5,4\right),\left(-5,-4,4\right),\left(4,-4,-5\right)\right\},\\
        A_3&=\left\{\left(-4,-4,5\right),\left(-4,5,-4\right),\left(5,-4,-4\right)\right\},\text{ and}\\
        A_4&=\left\{\left(4,4,5\right),\left(-5,4,-4\right),\left(4,-5,-4\right)\right\}.
    \end{align*}

Then their social choices are
\begin{align*}
    f(A_1)&=\left(-1,1,1\right),\\
    f(A_2)&=\left(1,-1,1\right),\\
    f(A_3)&=\left(-1,-1,-1\right),\text{ and}\\
    f(A_4)&=\left(1,1,-1\right).
\end{align*}

Under this construction, $\left(5,4,4\right)$ and $\left(-4,-5,4\right)$ prefer $f(A_2)$ over $f(A_1)$ because $d\left(\left(5,4,3\right),f(A_1)\right)=d\left(\left(-4,-5,4\right),f(A_1)\right) =3\sqrt{6}$, while $d\left(\left(5,4,4\right),f(A_2)\right)=d\left(\left(-4,-5,4\right),f(A_2)\right)=5\sqrt{2}<3\sqrt{6}$. By symmetry, two voters in each subpopulation prefer a different subpopulation, so there are arcs between each of these subpopulations. Therefore, the complete digraph with four vertices is a subgraph of the franchise graph of $N$.
\end{proof}

\begin{figure}
\centering
    \includegraphics[scale=0.5]{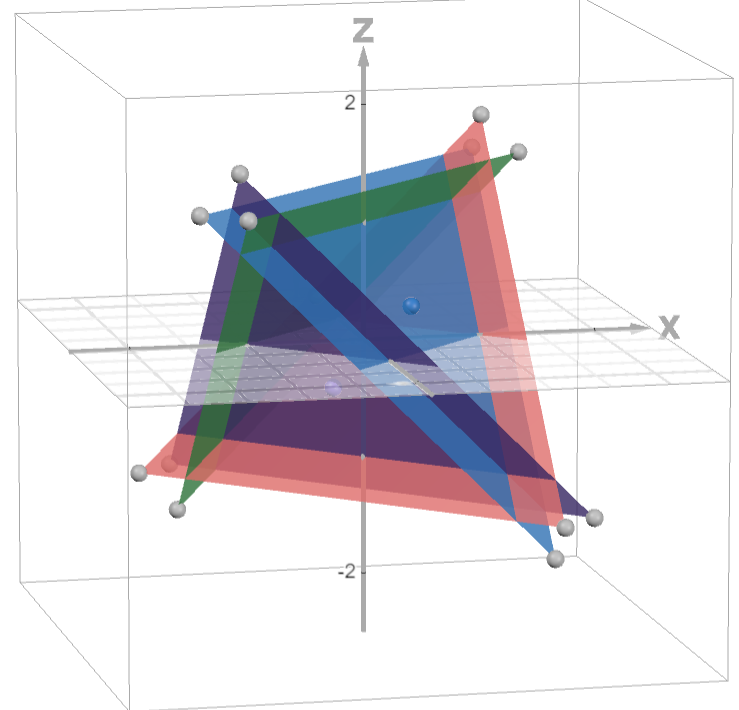}
    \label{fig:4CompleteGraph}
    \caption{The voters in each subpopulation are not in the convex hulls of any other subpopulation, so these subpopulations are Overton unstable.}
\end{figure}

Now, we describe a general construction for a population in $\mathbb{R}^n$ with an clique of size $n+1$.

\begin{thm}
    For $n\geq3$ and any rotationally symmetric and anonymous social choice function $f$, in $\mathbb{R}^n$, there exists a population with the complete $(n+1)$-digraph as a subgraph of its franchise graph.
\end{thm}

\begin{proof}
    We will construct a population $N$ with the complete $(n+1)$-graph as a subgraph of its franchise graph. Let $S_n=\{\mathbf{a_1},\mathbf{a_2},...,\mathbf{a_{n+1}}\}$ be the set of vertices of an $n$-dimensional regular simplex centered at the origin, and assume $||\mathbf{a_i}||=1$. Let $\mathbf{c_{-k}}=\frac{1}{n}\sum_{i\neq k}\mathbf{a_i}=\frac{1}{n}(\sum_{i=1}^{n+1}\mathbf{a_i}-\mathbf{a_k})=\frac{1}{n}(\mathbf{0}-\mathbf{a_k})=-\frac{1}{n}\mathbf{a_k}$, the center of the $(n-1)$-dimensional regular simplex with vertices $S_n\setminus\{\mathbf{a_k}\}$. Let $\mu>1$, and let $\mathbf{a_{i,k}}=\mu (\mathbf{a_i}-\mathbf{c_{-k}})+\mathbf{c_{-k}}=\mu(\mathbf{a_i}+\frac{1}{n}\mathbf{a_k})-\frac{1}{n}\mathbf{a_k}=\mu\mathbf{a_i}+\frac{1}{n}(\mu-1)\mathbf{a_k}$. Let $A_k=\{\mathbf{a_{1,k}},\mathbf{a_{2,k}},...,\mathbf{a_{n,k}}\}$, and let $N=\bigcup_{i=0}^{n+1}A_i$. Then, because $f$ is rotationally symmetric and anonymous, $f(A_k)=\mathbf{c_{-k}}$. We show that, for all $i,j,k\in\{1,2,...,n+1\}$, if $i\neq j$, $i\neq k$, and $j\neq k$, then $d^2(\mathbf{a_{i,k}},\mathbf{c_{-k}})>d^2(\mathbf{a_{i,k}},\mathbf{c_{-j}})$.
    \begin{eqnarray*}
        d^2(\mathbf{a_{i,k}},\mathbf{c_{-k}})&=&||\mathbf{a_{i,k}}-\mathbf{c_{-k}}||^2\\
        &=&\bigg|\bigg|\mu\mathbf{a_i}+\frac{1}{n}(\mu-1)\mathbf{a_k}-\biggl(-\frac{1}{n}\mathbf{a_k}\biggr)\bigg|\bigg|^2\\
        &=&\mu^2\biggl(\mathbf{a_i}+\frac{1}{n}\mathbf{a_k}\biggr)\cdot\biggl(\mathbf{a_i}+\frac{1}{n}\mathbf{a_k}\biggr)\\
        &=&\mu^2\biggl(\mathbf{a_i}\cdot\mathbf{a_i}+\frac{2}{n}\mathbf{a_i}\cdot\mathbf{a_k}+\frac{1}{n^2}\mathbf{a_k}\cdot\mathbf{a_k}\biggr)\\
        &=&\mu^2\biggl(1+\frac{2}{n}\biggl(-\frac{1}{n}\biggr)+\frac{1}{n^2}\biggr)\\
        &=&\mu^2\biggl(1-\frac{1}{n^2}\biggr).\\
        d^2(\mathbf{a_{i,k}},\mathbf{c_{-j}})&=&||\mathbf{a_{i,k}}-\mathbf{c_{-j}}||^2\\
        &=&\bigg|\bigg|\mu\mathbf{a_i}+\frac{1}{n}(\mu-1)\mathbf{a_k}-\biggl(-\frac{1}{n}\mathbf{a_j}\biggr)\bigg|\bigg|^2\\
        &=&\biggl(\mu\mathbf{a_i}+\frac{1}{n}(\mu-1)\mathbf{a_k}+\frac{1}{n}\mathbf{a_j}\biggr)\cdot\biggl(\mu\mathbf{a_i}+\frac{1}{n}(\mu-1)\mathbf{a_k}+\frac{1}{n}\mathbf{a_j}\biggr)\\
        &=&\mu^2\mathbf{a_i}\cdot\mathbf{a_i}+\frac{1}{n^2}(\mu-1)^2\mathbf{a_k}\cdot\mathbf{a_k}+\frac{1}{n^2}\mathbf{a_j}\cdot\mathbf{a_j}+\frac{2\mu}{n}(\mu-1)\mathbf{a_i}\cdot\mathbf{a_k} \\ & &+  \frac{2\mu}{n}\mathbf{a_i}\cdot\mathbf{a_j}+\frac{2}{n^2}(\mu-1)\mathbf{a_j}\cdot\mathbf{a_k}\\
        &=&\mu^2+\frac{1}{n^2}(\mu-1)^2+\frac{1}{n^2}-\frac{2\mu}{n^2}(\mu-1)-\frac{2\mu}{n^2}-\frac{2}{n^3}(\mu-1)\\
        &=&\mu^2\biggl(1-\frac{1}{n^2}\biggr)-\frac{2}{n^2}\biggl(1+\frac{1}{n}\biggr)(\mu-1)\\
        &<&\mu^2\biggl(1-\frac{1}{n^2}\biggr).\\
    \end{eqnarray*}

    Thus, $d^2(\mathbf{a_{i,k}},\mathbf{c_{-k}})>d^2(\mathbf{a_{i,k}},\mathbf{c_{-j}})$, so for all $j,k\in\{1,2,...,n+1\}$, if $j\neq k$, then $n-1$ of $n$ voters in subpopulation $A_k$ prefer $f(A_j)=c_{-j}$ to $f(A_k)=c_{-k}$, which for $n\geq3$ is a majority. Therefore, there is an arc from $A_k$ to $A_j$, and the subgraph with nodes $A_1$, $A_2$, ..., $A_{n+1}$ is the complete $(n+1)$-graph, which is a subgraph of the franchise graph of $N$.
\end{proof}

Thus, for at least any rotationally symmetric and anonymous social choice function and Euclidean embeddings, the franchise graph can include cycles, and for $\mathbb{R}^3$ and above, arbitrarily long cycles. Moreover, the number of cycles can grow with the embedding dimension, as the franchise graph can include cliques of sizes that grow with the embedding dimension. Thus, while franchise elections dynamics benefit from many fewer possible proposals compared to spatial voting, they remain dynamically rich. 

\section{Discussion}

As we have shown, while the median voter theorem guarantees franchise stability when there is only a single political axis, the size and shape of the electorate may be unstable if there is a second ideological axis. We have illustrated this instability for a population that uses the yolk center as their social choice function, showing the phenomenon where introducing as an iconoclast group a new ideological direction to a left-right polity can lead to the iconoclast group or extreme left or right voters becoming disenfranchised. Also as part of this phenomenon, we found that retaining a second ideological axis requires that the iconoclasts ally with an existing end of the political axis on which a majority lies, and that the iconoclasts will be able to remain in the electorate with a more iconoclastic stance if they ally with the side that is farther from the center. 

That retaining a new issue axis among a population traditionally embedded in a one-dimensional ideological space requires strategically associating positions on the new axis with stances on the pre-existing axis may be of particular relevance to scientific or activist communities, as these groups are often among the first to be aware of a new issue and, due to their reputation and reach, among the first to be able to add the new issue to the area of acceptable discourse. For such groups, we believe that, due to this phenomenon, they risk becoming irrelevant to mainstream discussions if they introduce new issues from positions which are neutral on the primary ideological axis. To avoid this, whether the optimal move is to ally with one side politically, as we have modeled, and accept that this will likely politicize the issue, or to remain neutral on the primary axis and rely on constituents of the Overton window not actually behaving as we have modeled, likely depends on the issue and the group's goals. While the former may be a safer option if the iconoclasts wish to remain in the Overton window above all else, the latter may be a better option in real situations because, while we assume that voters' ideal points remain static, actual scientists or activists may be able to convince people to move their ideal points.

The result that franchise instability requires multiple embedding dimensions also illustrates the reason that previous models of electoral enfranchisement which assumed single-dimensional political embeddings required additional modeling assumptions to capture the expansion of the franchise. As the median voter theorem suggests, any single-dimensional ideological axis is, at best, neutral to changes in the franchise, so the introduction of additional voters on the axis can only occur despite the enfranchised population's opinion. While these single-dimensional models can show the effects of outside influences changing the enfranchised population, they cannot capture the same situations our model does, as in, any situations where an enfranchised population chooses to change. However, our model would also have probable shortcomings in demonstrating the same situations as these single-dimensional models, for it is likely a debatable choice to attempt to model an elite class meddling with the electorate as voters on additional ideological axes. Whether such a choice could be appropriate would heavily depend on the modeled situation, its historical context, and the expected behavior of voters.

While we have shown that our model has more demonstrative potential than single-dimensional models in that it can show voluntary changes to the enfranchised population, we also have addressed the greater potential for instability that working in two or more dimensions begets. In corollary~\ref{cor:some_N_unstable}, we show that franchise instability is a fundamental element of spatial voting in Euclidean spaces of dimension two and above, as there are populations that are franchise, Overton, and disenfranchisement unstable for all social choice functions with the unanimity property, which certainly includes all reasonable social choice functions for our model. Additionally, our model provides some insights into franchise instability that remains over possibly infinite sequences of elections. In our investigation of franchise dynamics, we have shown that, in Euclidean spaces of dimension two and above, cycles and cliques may appear in the franchise graph. Thus, there exist populations with corresponding infinite sequences of winning alternative enfranchised populations that never reach a franchise stable population. As a consequence, even though the average center of the yolk may be an effective enough summary of the social choice of a population for our purposes, fundamental issues surrounding franchise instability inherent to multidimensional franchise selection limit its ability to describe the long-term social choice of a changing electorate.  

These results highlight the extent to which self-interested but myopic voting can destabilize the electoral composition. While we have largely discussed franchise elections in this paper as a simplified model for the Overton window in public discourse, with some additional assumptions, it may also describe franchise changes as the subject of actual elections. In particular, franchise elections may begin to model actual issues in political franchisement when voters are not only myopic, but when the franchised population is determined based on proxies to the ideological space, rather than actual placement on the ideological space. Namely, while it is conceivable that franchise changes can be enacted explicitly along an ideological axis (i.e. enfranchising only members of a specific political party), in non-totalitarian settings, it is more common for restrictions or modifications to the enfranchised population to occur based on properties that commonly correspond with the placement of a person's voter ideal point on the ideological axes, or ideological proxies. For instance, political structures that amplify rural populations, notably the U.S. Senate and Electoral College, are systems that elevate some voters, diminishing others, along a practical axis (rural-ness) with clear ideological implications. While elements of a person's identity such as group affiliations, professions, or life choices are likely to correlate with placement on some ideological axis, it is unlikely that every ideological issue has a convenient covariate against which to restrict the franchise. However, when such a proxy exists, questions about franchise stability may become relevant. In those cases, to the extent that preserving voting rights inside a democracy is important, myopic issue voting can be a dire threat when ideological space is multidimensional. In contrast, we did not include normative commitments to preserving the franchise, or model far sighted voters interested in the long run outcome of changing the franchise rules, but such stabilizing forces are likely vital to ensuring actual long-term, robust stability. Indeed, when issue polarization begins to place the social choice on individual issues as more important than commitments to political process or structure, that may signal the beginning of the dynamics studied of this paper, where portions of the electorate decide they would prefer other portions not to vote. While this may sound alarmist, if we include efforts to reduce voter turnout or to gerrymander political districts as functionally equivalent to disenfranchisement or partial disenfranchisement, then the dynamics of this paper are already occurring in the U.S. In particular, the growth and maintenance of gerrymandering reflects actions approved by the electorate under the representational rules established by that gerrymandering. It is possible that gerrymandering would not be supported by a majority of the electorate under the representational rules if there was no gerrymandering, leading to the sort of bistability or multi-stability exactly described by the above franchise elections.  

Thus, while we might hope that institutions and civically minded voters stand as a bulwark against mass disenfranchisement, our results suggest we should also be worried about the salience of issue politics. Namely, to the extent that extreme issue polarization can increase the salience of issues over normative commitments, ideological distributions with greater polarization can present a threat to the franchise. Aligned with the advice of many other commentators, we suggest that issue polarization entails innate risks to the franchise.

In contrast to political disenfranchisement, Overton disenfranchisement---as in, the exclusion of views from acceptable public discourse---likely has a more complicated valence, which may make our model more realistically relevant to these latter situations. While it is a common view that the negative impact of voter disenfranchisement on the integrity of democratic institutions outweighs its possible positive impacts, it is also a common view that television news shows ought not to broadcast and promote people with especially noxious or offensive views. This apparent paradox in common beliefs---the importance of a full political franchise and the obligation to, at least minimally, curate public discourse---is probably justifiable, but we will leave the particulars of its resolution to wiser thinkers. Nonetheless, this paradox suggests that the ideal Overton franchise is not maximal. Indeed, while cycling in the political franchise implies non-maximal enfranchisement, and thus presents a threat to the well-being of the democracy, cycling in the Overton franchise may benefit public discourse, as such cycling may allow the Overton window to more effectively adapt to changes in viewpoints or the addition of viewpoints to the population. Moreover, as opposed to political franchise, there may be many different areas of public discourse, each with different Overton windows. For instance, someone with large followings on social media platforms may fail to be published in the editorial section of a major newspaper, and a seasoned editorialist may fail to connect with audiences on social media. In this sense, the multi-stability of different Overton windows models that public discourse may have different venues, which each welcome and allow different viewpoints. An immediate area of application unmodeled in our paper would be in a population with Overton windows using different thresholds for their elections, as avenues for public discourse can vary greatly in the extent of disagreement they allow.

The observation that voters may wish to disenfranchise only themselves aligns with the questions of group formation in \cite{barbera2001voting,berga2004stability,bergantinos2017societies,bergantinos2018societies}. However, unlike the situations in those papers, our model does not allow voters to unilaterally disenfranchise themselves. In some settings, it may be appropriate to consider a more restrictive notion of stability than we have used, one which requires robustness to self-disenfranchisement. However, using such a notion of stability also opens the door to allowing more complicated voter behaviors that favorably move (or limit unfavorable movement of) the expected social choice. For instance, if a voter knows that their honest vote moves the average social choice away from them, then will they choose to eject themselves entirely, or will they truthfully vote on a subset of issues and otherwise abstain? Or will they choose to simply not be truthful at all? Indeed, if faculty meetings may suggest an answer, then we have empirical support for all three behaviors, even if actual voters seem to rarely possess the wisdom to understand when their presence hurts their own interests.

\section{Future Work}
Overall, we believe our framework of franchise elections has made new questions regarding voting games natural and approachable. Of those possible questions, there are several questions we currently conceive as most natural, as they are closely related to our results. 

Our proof of corollary~\ref{cor:some_N_unstable} showed that some specific populations must be franchise unstable for any social choice function, but this leaves the question: which populations without a core are franchise unstable? And in particular, are there efficient algorithms to determine whether a population is stable? While there have been significant improvements in the computational efficiency of calculating a population's yolk, and even more improvements in computing close relatives of the yolk, determining franchise stability remains daunting. Showing the franchise stability of any reasonably sized population would require somehow bounding the combinatorial explosion of possible competing social choice centers, as actually computing the yolk center of every competing population remains unreasonable. An efficient algorithm for finding stable populations would open up the questions in this paper to sampling disenfranchisement graphs, or walks on those graphs, allowing for a greater sense of what might be expected. Additionally, since franchise elections occur on a finite franchise graph and each voter has its own utility for each vertex in the franchise graph, then it may be possible for voters to associate each vertex in the franchise graph with a long-run utility, which would allow voters to escape the myopic assumption we have ascribed to them. 

The question raised in the discussion about multiple Overton windows with different voting thresholds seems likely to be immediately productive. Indeed, exploring different voting thresholds may also represent an exploration of a family of social norms that may help stabilize the franchise, though  section 3 serves as warning that neutral iconoclasts can be ejected for any threshold where they lack a veto. 

Lastly, on the question of social norms, families of appropriate norms not only likely already exist and have likely already have been acknowledged by scholars, but we believe incorporating these into the language of our model will be clarifying. This would especially be the case if the norms required for franchise stability exhibit trade-offs against other reasonable design goals for a democratic electoral system, as we suspect may be the case. 

\bibliographystyle{unsrt}  
\bibliography{references} 

\section{Appendix}
To setup the full generality for theorem~\ref{thm:total_dom_franchise}, consider a space $V$ and a population $N = \{x_1, x_2, \ldots, x_n \}$, where $x_i\in V$ for all $i= 1\ldots n.$ Further, suppose there is some symmetric utility function $u:V\times V \times V \to \{0,1\}$, such that $u(p_1,p_2,x) = 0$ is interpreted as a member at location $x$ preferring $p_1$ to $p_2$ and $u(p_1,p_2,x) = 1$ is the reverse, with a preference for $p_2$ over $p_1$, with symmetry requiring that $u(p_1,p_2,x) = 1- u(p_2,p_1,x) $. From the population and function $u$ we then define a majority election between candidate $p_1$ and $p_2$ in the standard way, namely if $\sum_i=1 ^n u(p_1,p_2,x) > \lceil n/2 \rceil$ we say that $p_2$ wins, if $\sum_i=1 ^n u(p_1,p_2,x) < \lceil n/2 \rceil$ we say that $p_1$ wins, and if neither of those occurs we declare it a tie.

Notice that this setting is significantly more general than the Euclidean setting, and more general than even any metric space embeddings, and is much closer to a generic social choice setting. The definition of a dominating set, and a total dominating set can safely be generalized to this setting, as can theorem~\ref{thm:total_dom_franchise} without alteration of its proof. 

\subsection{Statement on AI Use}
AI was used to help copy edit human written drafts of this manuscript.

\end{document}